\documentclass{svjour3}                     % onecolumn (standard   format)
\smartqed  % flush right qed marks, e.g. at end of proof
\usepackage{amsmath,amssymb,bm}
\usepackage{graphicx}
\usepackage{graphicx,graphics,color,epsfig}
\usepackage{mathptmx}      % use Times fonts if available on your TeX system
\usepackage{amsfonts}
\usepackage{amssymb}%
\usepackage[numbers,sort&compress,merge]{natbib}
 \journalname{Int. J. Theor. Phys.}

\begin{document}

\title{Bipartite Entanglement in Optomechanical Cavities driven by Squeezed Light}

%\subtitle{Bipartite Entanglement in Optomechanical cavities}

\titlerunning{Stationary Optomechanical Entanglement} 

\author{Smail Bougouffa and Mohannad Al-Hmoud}

\institute{ S. Bougouffa (Corresponding Author) \at  Physics Department, College of Science, Imam Mohammad ibn Saud Islamic University (IMSIU), P.O. Box 90950, Riyadh 11623, Saudi Arabia \\\email{sbougouffa@hotmail.com or sbougouffa@imamu.edu.sa}
\and  M. Al-Hmoud \at  Physics Department, College of Science, Imam Mohammad ibn Saud Islamic University (IMSIU), P.O. Box 90950, Riyadh 11623, Saudi Arabia\\
 \email{m\_hmoud@yahoo.com} }

\maketitle

\begin{abstract}
We investigate the stationary bipartite entanglement in a useful hybrid optomechanical system, which is constituted of two coupled-cavity optomechanics through a photon hopping process and both are driven by squeezed light. The transfer of correlations from an entangled light source to optomechanical cavities is explored. It is found that the generation of bipartite entanglement and entanglement transfer depend strongly on photon hopping strength and the matching of the input squeezed modes to the cavity modes. It is revealed that the generated stationary bipartite entanglement due to squeezed light that drives the cavities is robust against the thermal fluctuations. The fidelity of a coherent state of the optical modes is explored and it is shown that it offered interesting conditions on the stability of the system, which are the same for entanglement generation.



\end{abstract}

%\linenumbers

\section{Introduction}

Hybrid optomechanical systems are a combination of mechanical resonators, atoms, and optical cavities in different ways  \cite{Aspelmeyer2008,Blencowe2004,Genes2009,Aspelmeyer2010,Clerk2014} and extensively employed for a huge variety of applications such as detection of gravitational waves \cite{Vitali2007}, precision measurements of small displacement and force \cite{LIGO2007}, the light readout and storage \cite{Oo2013} and information processing, and quantum communication \cite{Liu2013,Yan2015,Yan2019}

It is well known that entanglement is a kind of quantum correlations and is at the heart of quantum information processing \cite{Braunstein}. Further, entanglement is recognized as one of the distinctions between classical and quantum systems \cite{Vedral2006,Sun2017}. Now, the ways of generation of entanglement among microscopic entities are developed and mastered \cite{Adessob,Vidal}. 

The problem of transferring quantum correlations from an entangled light source to initially separable or coupled cavity optomechanics is at the heart of concern with storage of quantum correlations in quantum memories for continuous variable quantum information processing and quantum-limited displacement measurements \cite{Lee2005,Bougouffa2012,Bougouffa2013b,Sete:14,sete2015high,Yan2015}.

Recently, there has been substantial attention in examining entanglement in mesoscopic systems \cite{Zhou2011,nunnenkamp2011single,Purdy2013a,Bai2016,Bougouffa2016,Liang2019} and nanomechanical oscillators have become the key resources for the exploration of quantum mechanical characters at mesoscopic scales \cite{Ge2013a,Ge2015,Ge2015a,Si2017,Asiri2018}. 

 Indeed, the generation of entanglement between nanomechanical oscillators from one side and between nanomechanical resonator and optical mode on the other side, has been examined in various ways: such as entangling a nanomechanical resonator and a superconducting microwave cavity \cite{vitali2007entangling}, entangling of a micromechanical resonator with output optical fields \cite{Genesc}, entangling of two mirrors of two coupled optical cavities among photon hopping process  \cite{liao2014entangling}, entangling optical and microwave cavity modes by means of a nanomechanical resonator \cite{Barzanjeh,Barzanjeh2018}, entangling two dielectric membranes suspended inside a cavity \cite{Hartmann2008}, entangling nanomechanical oscillators in a ring cavity by feeding squeezed light \cite{huang2009entangling}, entangling mechanical motion with microwave fields \cite{Palomaki2013}, entangling light to matter in ultra-strong coupling regime \cite{FriskKockum2019} and entangling two micromechanical oscillators \cite{Riedinger2018a}.

The optomechanical interaction is used to induce entanglement between the mechanical motion and the light beam. Consequently, the generated states can be described as a “Schrödinger cat” type state, where a “microscopic” degree of freedom (the optical cavity mode) is entangled with a “macroscopic” (or mesoscopic) degree of freedom, the vibrating mirror. 
The generated entanglement offers the canonical implementation for quantum information protocols involving continuous variables systems \cite{Hofer2011}. Accordingly, one can teleport an arbitrary input state of the light field onto a mechanical oscillator \cite{Pirandola2003,mancini2003scheme, Asjad2016}. Optomechanical devices, therefore, offer an effective supplement to the huge range of physical systems that are being investigated for quantum information processing. 

In addition, some schemes have been proposed to establish a transfer of entanglement from entangled light to separated optomechanical cavities \cite{PhysRevA.85.043824,ElQars2017,Kronwald,Yousif2014}.

In this work, we conceptually investigate an interesting technique to preserve and affect entanglement between the mechanical resonator and intracavity mode in hybrid cavity optomechanics.  We suggest a scheme that allows to follow the physical origin of quantum correlations and to study the transfer of correlation form entangled light to two optomechanical systems, which are separated or coupled among the photon hopping process. In this scheme, the entanglement between the movable mirror and its intracavity mode can be affected once a photon hopping process is established between the two cavity optomechanics. Moreover, when the two cavities are driven by a squeezed light, entanglement between the optical modes can be generated and the intra-entanglement is preserved and affected, i.e., a transfer of entanglement between bipartite subsystems takes place.

The paper is structured as follows: In Section \ref{sec2}, we present the model of a hybrid optomechanical cavity, which is driven by squeezed light and derive the Hamiltonian for the proposed scheme. In section \ref{sec3}, the effective quantum Langevin equations are derived in the rotating wave approximation. We then employ the linearization technique to the equations of motion and get a system of coupled differential equations for the fluctuation operators, which can be solved in the steady-state regime. In Section \ref{sec4}, we study the covariant matrix of the system and use the logarithmic negativity to quantify the degree of bipartite entanglement of subsystems. In Section \ref{sec5}, we discuss the obtained results of entanglement between different modes and explore the stability conditions within the experiment parameter ranges. Finally, a conclusion is given in Section \ref{sec6}.

\section{Model and Hamiltonian}\label{sec2}
\begin{figure}[ht]
 \center{\includegraphics[width=0.60\linewidth,height=0.40\linewidth]{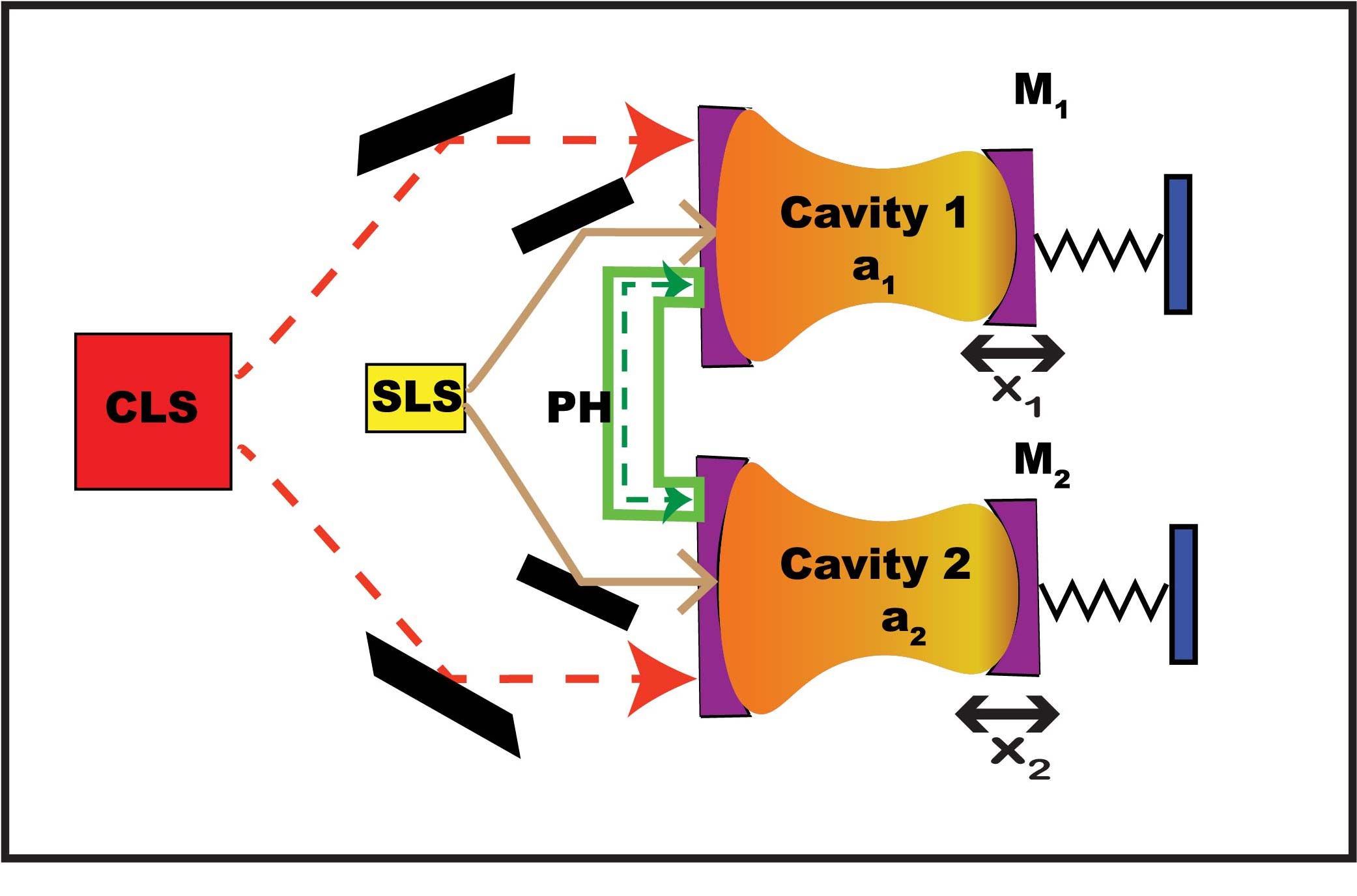}}
\caption{(Color online)The coupling configuration of the output field of a squeezed light source (SLS) to a pair
of single-mode optomechanics cavities 1 and 2, which they are coupled with photon hopping process (PH) and driven by a coherent light source (CLS). }\label{Fig1}
\end{figure}

We consider a hybrid optomechanics system consisting of two optomechanical cavities which are coupled to each other via a photon hopping (PH) process, both fixed sides are exposed to the output field of a squeezed light source (SLS) and a coherent light source (CLS). The other side of the two cavities is moving mirrors as sketched in Fig.\ref{Fig1}. 

The Hamiltonian describing the unitary dynamics of the scheme, through the rotating wave approximation in an appropriate observation setting, reads  ($\hbar=1$)
\begin{eqnarray}\label{eq1}
 \hat{H}&=&\sum_{j=1,2}\Big[\Delta_{0j} a_{j}^\dagger a_{j}
         + \frac{1}{2}\omega_{mj}(p_{j}^2+ q_{j}^2)
         - g_{j} a_{j}^\dagger a_{j} q_{j}
         +i(E_j a_{j}^\dagger-E_j^{*} a_{j} )\Big] \nonumber \\&{}&-\xi \big( a_{1}^\dagger a_{2}+a_{2}^\dagger a_{1}\big).
\end{eqnarray}
The first term represents the energy of each cavity field with $a_{j}^\dagger$ and $a_{j}$  ($j=1, 2$) are their increasing and decreasing operators, respectively, where $([a_{j}^\dagger,a_{k}]=-\delta_{jk})$ and $\Delta_{0j}=\omega_j-\omega_L$ is the detuning of cavity fields. The second term describes the energy of the two mechanical resonators with a frequency of $\omega_{mj}$.  $q_{j}$ and $p_{j}$ are their dimensionless position and momentum operators, which satisfy the commutation relation $([q_{j},p_{k} ]=i \delta_{jk})$. The third term is the radiation pressure interaction with coupling strength $g_{j} =(\omega_{cj}/L_{j})\sqrt{ \hbar/(m_{j} \omega_{mj})}$, where $L_{j}$ is the rest length of each optomechanical cavity and $m_{j}$ is the mass of each mechanical resonator, respectively. The fourth term shows the driving field with frequency $\omega_{Lj}$ and the amplitude $E_j$ is expressed in terms of input field power $P_{j}$ by $|E_j|=\sqrt{\frac{2P_{j}\kappa_{j}}{\hbar\omega_{Lj}}}$, where $\kappa_j$ is the decay rate of each cavity field. The last term defines the photon hopping coupling between the two cavity modes with a strength of $\xi$. It is remarkable to state here that some preceding investigations have studied multi-cavity optomechanical systems with one mechanical oscillator \cite{liao2014entangling,Guo2014} or two mechanical resonators in the deep-resolved-sideband regime \cite{liao2015enhancement,clerk2008back}. In addition,  the two optomechanical cavities are pumped by two-mode squeezed light, of equal frequencies $\omega_s$. The squeezed field from a degenerate parametric oscillator (DPO) is characterized by the photon number $N(\omega_k)$, the two-photon correlations $M(\omega_k)$, and the squeezed field phase $\phi_s$. The squeezing properties are given by \cite{Tanas2002,messikh2004effect}
\begin{eqnarray}
N(\omega_{k}) &= \frac{\lambda^{2} - \mu^{2}}{4}\left[\frac{1}{(\omega_{k} -\omega_{s})^{2} + \mu^{2}} -\frac{1}{(\omega_{k} -\omega_{s})^{2} + \lambda^{2}}\right], \label{e2}\\
M(\omega_{k}) &= \frac{\lambda^{2} - \mu^{2}}{4}\left[\frac{1}{(\omega_{k} -\omega_{s})^{2} + \mu^{2}} +\frac{1}{(\omega_{k} -\omega_{s})^{2} + \lambda^{2}}\right], \label{e3}
\end{eqnarray}
where $\mu =\frac{1}{2}\kappa-\varepsilon$, $\lambda =\frac{1}{2}\kappa+\varepsilon$, with $\kappa$ the damping constant of the DPO, and $\varepsilon$ is its amplification parameter, proportional to the amplitude of the pumping field. The parameters $\kappa$ and $\varepsilon$ can be varied to produce squeezed fields of different bandwidths and intensities, and strong squeezing occurs over the bandwidth $\mu$ when $|\varepsilon|\rightarrow \kappa/2 $.
In the following, we are concerned in the dynamics of the system that can be described by the quantum Langevin equations.

\section{Quantum Langevin Equations}\label{sec3}

The analyses of the system dynamics are determined by the fluctuation-dissipation processes affecting both the cavity and the mechanical mode. Using the Hamiltonian (\ref{eq1}) and taking into account dissipation processes, one can obtain the following set of nonlinear quantum Langevin equations, written in the interaction picture for $\hbar\omega_{L j} a_{j}^{\dag}a_{j}$,
\begin{eqnarray}\label{eq2}
    \dot{q_{1}}&=&\omega_{m1} \ p_{1}, \nonumber\\
     \dot{q_{2}}&=&\omega_{m2} \ p_{2}, \nonumber\\
    \dot{p_{1}}&=&-\omega_{m1} \ q_{1} + g_{1} {a_{1}}^{\dag}a_{1}-\gamma_{m1} \ p_{1}+\sqrt{2\gamma_{1}}b^{in}_{1}, \nonumber\\
        \dot{p_{2}}&=&-\omega_{m2} \ q_{2} + g_{2} {a_{2}}^{\dag}a_{2}-\gamma_{m2} \ p_{2}+\sqrt{2\gamma_{2}}b^{in}_{2}, \nonumber\\
    \dot{a_{1}}&=&-(\kappa_{1}+i \Delta_{01})a_{1} +i g_{1} a_{1} q_{1} +i \xi a_{2}+E_1 +\sqrt{2\kappa_{1}} a^{in}_{1},\nonumber\\
     \dot{a_{2}}&=&-(\kappa_{2}+i \Delta_{02})a_{2} +i g_{2} a_{2} q_{2} +i \xi a_{1}+E_2 +\sqrt{2\kappa_{2}} a^{in}_{2},
\end{eqnarray}
where  $\gamma_{mj}=\omega_{mj}/Q_{mj}$  is the damping rate of the mechanical mode $mj$. Further, we have also comprised the noises of the input modes $b_j^{in}$ and $a_j^{in}$ appearing from the coupling of the modes to their surrounding environments. Now, $a^{in}_{j}$ is the squeezed vacuum operator with the following correlation functions \cite{Gao,Gardiner1986a,Carmichael1987a,Carmichael1987d,Parkins1990,dalton1999atoms,huang2009entangling}:

\begin{eqnarray}\label{eq3}
\langle a_j^{in}(t) {a_{j}^{in}}^{\dagger}(t')\rangle &=&(N_{j}+1)\delta(t-t'),\nonumber\\
\langle {a_j^{in}}^{\dagger}(t) a_{j}^{in}(t')\rangle &=&N_{j}\delta(t-t'), \nonumber\\
\langle a_j^{in}(t) a_{j'}^{in}(t')\rangle &=&M_{jj'}\delta(t-t'),\quad j\ne j' ,\nonumber\\
\langle {a_j^{in}}^{\dagger}(t) {a_{j'}^{in}}^{\dagger}(t')\rangle&=&M_{jj'}^{*} \delta(t-t'),\quad j\ne j' ,
\end{eqnarray}
where we assume that the squeezed light in a broad-band squeezed vacuum state centered about the frequency $\omega_L$, and that the bandwidth of the squeezing is sufficiently broad that the squeezed vacuum appears as $\delta$-correlated squeezed white noise to the cavities. In addition, without loss of generality, the parameters $N_j$ and $M_{jj'}$ are now assumed independent of the frequency. On the other hand,  the factor $|M_{jj'}|$ may belong to one of the two separate regions:
\begin{equation}
\label{ eq4}
|M_{jj'}| < N_{j}  \quad \textrm{or} \quad N_{j} < |M_{jj'}| \leq \sqrt{N_{j}(N_{j}+1)}.
\end{equation}
The region of $|M_{jj'}| < N_{j}$ corresponds to the classically squeezed field in the way that fluctuations in one of the quadratures of the field amplitudes are reduced but not below the shot-noise level. While, in the second region $N_{j} <|M_{jj'}| \leq \sqrt{N_{j}(N_{j} + 1)}$,  the field is then a quantum squeezed field in the way that the fluctuations of one of the quadratures are repressed below the shot-noise level. The case  $|M_{jj'}| = \sqrt{N_{j}(N_{j} + 1)}$ matches to maximal correlations, an ideal squeezed field. Consequently, the two bounds $|M_{jj'}|= N_{j}$ and $|M_{jj'}|=\sqrt{N_{j}(N_{j}+1)}$ fall into the quantum correlations of the squeezed field. 

For the statistics of the input mechanical modes, we suppose that the mechanical quality factor  $Q_{m_{j}} \gg 1$. In this limit, the Brownian noise $b_j^{in}$ can be approximated as a Markovian process. Thus, the phonon modes in thermal vacuum states are characterized by the correlation functions. \cite{giovannetti2001phase,Rehaily2017,Gardiner,Benguria,Zhang}
\begin{eqnarray}\label{eq5}
\big\langle b_j^{in}(t) {b_j^{in}}^{\dagger}(t')\big\rangle &=&\ (\bar{n_{j}}+1)\delta(t-t'),\nonumber\\
\big\langle {b_j^{in}}^{\dagger}(t) b_j^{in}(t')\big\rangle &=&\ \bar{n_{j}}\delta(t-t')
\end{eqnarray}
where $\bar{n_{j}} = (e^{\hbar \omega_{mj}/k_{B} T} - 1)^{-1} $ is the mean number of thermal phonons at the frequency of the mechanical resonator $j$, $k_b$ is the Boltzmann constant, and T is the temperature of the environment.

We can realize from the system (\ref{eq2}) that the modes $a^{R}_{1}$ and $a^{R}_{2}$ are directly coupled to each other with the strength $\xi$, and are also indirectly coupled to each other across the coupling to the corresponding  vibrating mode with strengths $g_1$ and $g_2$, respectively. This coupling arrangement does not exhibit a closed procedure and hence the dynamics of the system cannot show phase-dependent effects \cite{Sun2017}.

For enough driving power and high-finesse cavities, the system can be described by a semiclassical steady state with the cavity mode amplitude $a^s_{j}$  $(|a^s_j|\gg  1)$, and a new equilibrium position for the oscillators, displaced by $q^{s}_{j}$. The parameters $p^{s}_{j}$, $q^{s}_{j}$ and  $a^{s}_{j}$ are the solutions of the nonlinear algebraic equations obtained by factorizing Eqs. (\ref{eq2}) and setting the left-hand sides to zero, yielding
\begin{eqnarray}\label{eq6}
p^s_{j}&=&0, \quad q^s_{j}=\frac{g_{j} \mid a^s_{j}\mid^{2}}{\omega_{mj}}; \quad j=1,2,\nonumber\\
a^s_{1}&=&\frac{\alpha_2E_1+i\xi E_2}{\alpha_1\alpha_2+\xi^2},\nonumber\\
a^s_{2}&=&\frac{\alpha_1E_2+i\xi E_1}{\alpha_1\alpha_2+\xi^2},
\end{eqnarray}
where $\alpha_j=\kappa_j+i\Delta_j, (j=1,2).$
The last two equations of (\ref{eq6}) are indeed nonlinear equations providing the steady intracavity field amplitude $a^s_{j}$, as the effective cavity detuning $\Delta_{j}$, including radiation pressure effects, which is given by $\Delta_j=\Delta_{0j}-\frac{g_j^{2}}{\omega_{mj}}|a^s_{j}|^2$. The parameter regime suitable for producing optomechanical entanglement is that with a very large input power $P$, i.e., as $|a^s_{j}|\gg 1$. Also, The nonlinearity exhibits that the steady intracavity field amplitude can expose variability behavior for a certain parameter regime. On the other hand, the two equations are coupled to each other by the factor $\xi$, which is due to the photon hopping process between the two cavity modes. 

Now, the equations of motion (\ref{eq2}) may be solved by the linearization approach \cite{Braunstein2012}. In this approach, we assume  that the operators are changed by small fluctuation from their steady state solutions  $a_j^R = a^{s}_{j}+\delta a_j , q_j = q^{s}_{j}+\delta q_j$ and $ p_j = p^{s}_{j}+\delta p_j$.
Keeping the linear terms only, the equations of motion for the fluctuation parts of the operators can be obtained as

\begin{eqnarray}\label{eq8}
    \delta\dot{q_{1}}&=&\omega_{m1} \ \delta p_{1}, \nonumber\\
    \delta\dot{q_{2}}&=&\omega_{m2} \ \delta p_{2}, \nonumber\\
    \delta\dot{p_{1}}&=&-\omega_{m1} \ \delta q_{1}-\gamma_{m1} \ \delta p_{1} + G_{1} \delta X_{1} +\sqrt{2\gamma_{1}}b^{in}_{1}, \nonumber\\
    \delta\dot{p_{2}}&=&-\omega_{m2} \ \delta q_{2}-\gamma_{m2} \ \delta p_{2} + G_{2} \delta X_{2} +\sqrt{2\gamma_{2}}b^{in}_{2}, \nonumber\\
    \delta\dot{X_{1}}&=&-\kappa_{1} \delta X_{1} - \Delta_{1} \delta Y_{1}-\xi \delta Y_{2} + \sqrt{2\kappa_{1}}\delta X_{1}^{in},      \nonumber\\
     \delta\dot{X_{2}}&=&-\kappa_{2} \delta X_{2} - \Delta_{2} \delta Y_{2}-\xi \delta Y_{1} + \sqrt{2\kappa_{2}}\delta X_{2}^{in},      \nonumber\\
    \delta\dot{Y_{1}}&=&-\kappa_{1} \delta Y_{1} + \Delta_{1} \delta X_{1} +\xi \delta X_{2}+ G_{1} \delta q_{1} + \sqrt{2\kappa_{1}}\delta Y_{1}^{in},\nonumber\\
     \delta\dot{Y_{2}}&=&-\kappa_{2} \delta Y_{2} + \Delta_{2} \delta X_{2} +\xi \delta X_{1}+ G_{2} \delta q_{2} + \sqrt{2\kappa_{2}}\delta Y_{2}^{in},
\end{eqnarray}
where $G_{j} =\sqrt{2}g_{j} a^s_{j}$ is the effective optomechanical coupling strength of the mode $j$ to the mechanical mode and $a^s_{j}$ is assumed to be real. Where we have used the quadratures of the cavity mode,
\begin{equation}\label{eq9}
   \delta X_{j} = \frac{\delta a_{j} +\delta a_{j}^\dag}{\sqrt{2}},\quad \delta Y_{j} = \frac{\delta a_{j} -\delta a_{j}^\dag}{i \sqrt{2}},
\end{equation}
and the analogous hermitian input quadrature noises
\begin{equation}\label{eq10}
    \delta X_{j}^{in} = \frac{a_{j}^{in} + {a_j^{in}}^{\dag}}{\sqrt{2}}, \quad    \delta Y_{j}^{in} = \frac{ a_{j}^{in} - {a_j^{in}}^{\dag}}{i \sqrt{2}}
\end{equation}

In following, we are interested in the exploration of the different bipartite entanglement in the regime where the stability of the multipartite system is realized.

\section{Bipartite Steady-State Entanglement }\label{sec4}

The intracavity optical and mechanical mode compose a bipartite continuous variable (CV) system.  We will examine the kind of linear quantum correlations between field modes and mechanical modes by considering the steady state of the correlation matrix of quantum fluctuations in this multipartite system.  As the set equations (\ref{eq8}) are linear and the noise operators have assumed to be in Gaussian state with zero-mean Gaussian state, thus the system is completely characterized by the corresponding symmetrize covariance matrix (CM), which can be read as
\begin{equation}\label{eq11}
\nu_{lm}=\frac{\langle \mu_{l}(\infty) \mu_{m}(\infty)+\mu_{m}(\infty)\mu_{l}(\infty)\rangle}{2},
\end{equation}
where $\mu_{l}(\infty)$ is the steady-state value of the $l^{th}$ component of the vector of quadrature fluctuations
\begin{equation}\label{eq12}
\boldsymbol{\mu}(t)=(\delta q_{1}(t),\delta p_{1}(t),\delta X_{1}(t),\delta Y_{1}(t),\delta q_{2}(t),\delta p_{2}(t),\delta X_{2}(t),\delta Y_{2}(t))^{T}.
\end{equation}
The time evolution of its components can be amended in proper structure as
\begin{equation}\label{eq13}
    {\boldsymbol{ \dot\mu}}(t)=\mathbf{A}\boldsymbol{ \mu}(t)+ \mathbf{ C}(t).
\end{equation}
Where $\mathbf{A}$ is the drift matrix,

\begin{equation}\label{eq14}
\mathbf{A}=
\left(
 \begin{array}{cccccccc}
 0            & \omega_{m1}   & 0            & 0               & 0              & 0                & 0                  & 0              \\
  -\omega_{m1} & -\gamma_{m1}      & G_{1}               & 0              & 0            & 0                  & 0          & 0 \\
  0            & 0             & -\kappa_{1}            & \Delta_{1}     & 0              & 0                & 0                  & -\xi              \\
 G_{1}             & 0             & -\Delta_{1} & -\kappa_{1}           & 0                & 0                  & \xi          & 0 \\                                                                0            & 0             & 0            & 0               & 0 & \omega_{m2}  & 0              & 0           \\
  0       & 0             & 0            & 0               & -\omega_{m2}    & -\gamma_{m2}& G_{2}            & 0              \\
  0            & 0             & 0            & -\xi               & 0              & 0             & -\kappa_{2}  & \Delta_{2}   \\
  0            & 0             & \xi        & 0               & G_{2}             & 0                & -\Delta_{2}      & -\kappa_{2}
  \end{array}
\right),
\end{equation}
and $\mathbf{C}(t)$ is the vector of noises,
\begin{equation}\label{eq15}
\mathbf{C}=(0,\sqrt{2\gamma_{1}}b_{1}^{in}(t),\sqrt{2\kappa_{1}}X_{1}^{in}(t),\sqrt{2\kappa_{1}}Y_{1}^{in}(t),0, \sqrt{2\gamma_{2}}b_{2}^{in}(t),\sqrt{2\kappa_{2}}X_{2}^{in}(t),\sqrt{2\kappa_{2}}Y_{2}^{in}(t))^{T}.
\end{equation}
In this case of no hopping photon process, we can get the shape of the two independent optomechanical cavities, which are characterized by diagonal blocks of the drift matrix. Additionally, the correlation is localized in each intracavity. Nevertheless, one can see from the diffusion matrix that the cavity modes decay to a common reservoir. So, the entanglement can be transferred between the two independent cavity modes. The entanglement transfer depends on the kind of environment \cite{Bougouffa2016, Aloufi2015, Bougouffa2013, Bougouffa2013b, Bougouffa2012}. Once the photon hopping process is included, the non-diagonal blocks of the covariance matrix become not null. Then, the bipartite entanglement is affected and preserved without the coupling to the environment.  In the following, we will investigate these features in details.

The steady-state CM can be determined by solving the Lyapunov equation
\begin{equation}\label{eq16}
    \mathbf{AW}+\mathbf{WA^T}=-\mathcal{Q},
\end{equation}
here $\mathcal{Q}$ represents the diffusion matrix, which is determined by the noise correlation functions,

%\scriptsize
\begin{footnotesize}
\begin{equation}\label{eq17}
\mathcal{Q}=
\left(
 \begin{array}{cccccccc}
  0  & 0   & 0     & 0      & 0     & 0     & 0          & 0      \\
  0  &\gamma_{m1}(2\bar{n}+1)& 0   & 0     & 0    & 0  & 0    & 0 \\
  0  & 0   &\kappa_{1}(2N+1) & 0  & 0  & 0  & 2 \kappa M  & 0 \\
  0  & 0   &0 & \kappa_{1}(2N+1)    & 0  & 0  & 0   & -2 \kappa M  \\ 
   0  & 0   &0    & 0      & 0 &0  & 0   & 0           \\
  0  & 0   &0  & 0  & 0  & \gamma_{m2}(2\bar{n}+1)& 0   & 0     \\
  0  & 0   &2 \kappa M  & 0  & 0  & 0  & \kappa_{2}(2N+1)  & 0   \\
  0  & 0   &0  & -2 \kappa M & 0  & 0  &0      & \kappa_{2}(2N+1)
 \end{array}
 \right),
\end{equation}
\end{footnotesize}
where $\kappa=\sqrt{\kappa_1\kappa_2}$. For simplicity, we assume that $N_1=N_2=N$, $M_1=M_2=M$ and $\overline{n_1}=\overline{n_2}=\bar{n}$.

The CM permits the estimation of  the stationary entanglement between different bipartite subsystems. Thus, to evaluate the pairwise entanglement, we export from the $(8 \times 8)$ covariance matrix $W$ a $(4 \times 4)$ submatrix $W_R$. This technique gives rise to four such cases of the sub-matrix $W_R$: (i) If the indices $i$ and $j$ for the element $w_{ij}$ are kept to the set $\big\{1, 2, 3, 4\big\}$, the sub-matrix $W_R=[w_{ij}]$ is produced by the first four rows and columns of $W$ and match to the covariance between the first intracavity photon-phonon coupling. Correspondingly, (ii) if the indices run over $\big\{5, 6, 7, 8\big\}$, $W_R$ is the covariance matrix of the second intracavity photon-phonon interaction. (iii) If the indices run over $\big\{1, 2, 5, 6\big\}$, $W_R$ labels the covariance between the two mechanical resonator modes. (iv) Lastly, if the indices run over $\big\{3, 4, 7, 8\big\}$, $W_R$ denotes the covariance matrix between the cavity modes.

Consequently, the logarithmic negativity $E_N$  can be implemented here as a good entanglement measure to quantify the entanglement of the two subsystems. The logarithmic negativity is expressed by \cite{Plenio,Adesso,Vidal}

\begin{equation}\label{eq18}
    E_N= \max\Big[0,-\ln(2\vartheta^{-})\Big].
\end{equation}
Here, $\vartheta^{-}$ is the smallest symplectic eigenvalue of partial transpose $W_R$ matrix and is given by
\begin{equation}\label{eq19}
    \vartheta^{-}= \frac{1}{\sqrt{2}}\big(\chi(W_R)-[\chi(W_R)^2-4\det(W_R)]^{1/2}\big)^{1/2},
\end{equation}
and $\chi(W_R)\equiv \det W_{1}+\det W_{2}-2\det W_{c}$ ,with $W_{1}, W_{2}$ and $W_{c}$ being 2$\times$ 2 block matrices
\begin{equation}\label{eq20}
W_R =
\left(
\begin{array}{ cc}
      W_{1} & W_{c}    \\
      W_{c}^{T} & W_{2}  
\end{array}
\right),
\end{equation}
Within this framework, we are now prepared to investigate the distribution of entanglement between different subsystems of the proposed scheme. 

Before continuing further, we mention here that several similar schemes have been investigated and optimal stationary entanglement between different subsystems is reached in diverse regimes \cite{liao2014entangling,PhysRevA.84.053817, Sete:14,Yousif2014, Ludwig, Rehaily2017, Zhang, Sete2015, Setea, Bai2016}. Then, the current investigation permits to present and compare our obtained results with the suggested scheme. 

On the other hand, we briefly comment the possible setups and the parameters of the suggested model could be established with the topical experiments. An appropriate possible realization of the suggested scheme could be, for example, the experimental system with two Fabry–Perot optical cavities and whispering cavities \cite{Guo2014}. The parameter regime very close to the current experimental results \cite{Guo2014, Groeblacher2009b, Schliesser2009, Fainsteina, Kleckner, Gigan2006,Groeblacher2008,Heidmann2008, Arcizet2006,Arcizet2009,Han2013a,Ockeloen-Korppi2018a,Pirkkalainen2013}, is given as $\big(L_j \simeq 1 \;\emph{mm}, m_j \simeq10\;\emph{ng}, \omega_{mj} /2\pi = 10\;\emph{MHz}, \gamma_{mj} /2\pi \simeq 100\;\emph{Hz}, \kappa_j /2\pi \simeq 5 \sim 15\; \emph{MHz}, T = 0.6 \sim 20\;\emph{K}, P = 50\;\emph{mW}, \lambda_j = 1064\;\emph{nm}\big). $

We point out that the processed theoretical results will be explicitly presented for the achievable parameters chosen from the experiments as mentioned above.  For simplicity, without loss of generality, we suppose that the two cavities are identical and we choose the same parameters for the two mechanical resonators and the two optical modes,  i.e; $\big(\omega_{m1}=\omega_{m2}=\omega_m, \kappa_1=\kappa_2=\kappa, \gamma_{m1}=\gamma_{m2}=\gamma_m, T_1=T_2=T, g_1=g_2=g, \Delta_{1}=\Delta_{2}=\Delta\big)$.

\begin{figure}[ht]
\center{
\includegraphics[width=0.4\linewidth,height=0.35\linewidth]{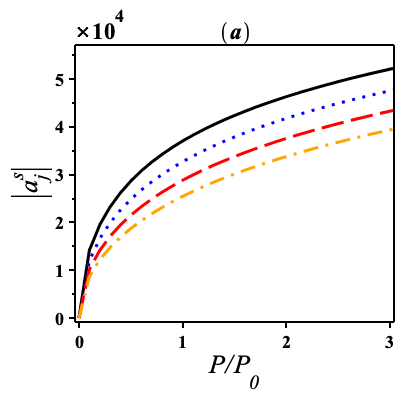}
\includegraphics[width=0.4\linewidth,height=0.35\linewidth]{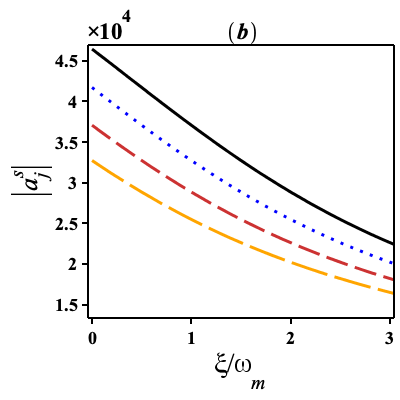}}
\caption{(Color online)The variation of the steady-state amplitudes $|a^s_{1}|=|a^s_{2}|$ , (a) versus the driving power $P/P_0$, (b) versus the photon hopping strength,  for different values of the detuning of cavity field.   $(\textrm{black solid line}:\Delta_{0}/\omega_m=0 ),(\textrm{blue dotted line}:\Delta_{0}/\omega_m=0 .5), (\textrm{red dashed line},\Delta_{0}/\omega_m=1), (\textrm{green dash dotted line},\Delta_{0}/\omega_m=1.5) $,  where $P_0=35 \emph{mW}$ . For (a) $\xi=\omega_m$ and for (b) $P/P_0=1$. The parameters are chosen to be ($L=1\; \emph{mm}, m_{j} = 5\; \emph{ng} , \omega_{m}/2\pi = 10\;  \emph{MHz}, \gamma_{m}/2\pi =100 \; \emph{Hz}, \kappa/2\pi = 14\;  \emph{MHz} , \lambda =810\;  \emph{nm}, \Delta=\omega_m$). }\label{Fig2}
\end{figure}

Now, We explore the stationary entanglement between different bipartite of the system and the fidelity of the two optical modes. We assume that the system reaches the steady state only if it is stable and this is fulfilled by the Routh-Hurwitz criteria. On the other hand, the amplitudes of the cavity and mechanical modes satisfy the conditions $|a^{s}_{j}|\gg1$ for the suitable regime parameters as shown in Fig. \ref{Fig2} and the amplitudes increase with increasing the driving power. For example, at the point of the driving power $\big(P\approx 35$ \emph{mW}, $\xi/\omega_m\approx 1,  |a^s_{j}|\approx 2.5\times 10^4 \gg 1\big)$, then these plots confirm the assumption of the linearization approach. Furthermore, the appropriate choice of the parameters can produce bipartite entanglement with the suggested scheme.

\section{ Results and Discussion}\label{sec5}
We now investigate the effects of the different introduced aspects, i. e.; photon hopping process and squeezed light, on the optomechanical entanglement and fidelity within the experiment parameter regime, which is suitable for producing bipartite entanglement. 

\begin{figure}[h]
\hspace*{2cm}\textbf{(a)} \hspace*{3.5cm}\textbf{(b)}\hspace*{3.5cm}\textbf{(c)}\\
\includegraphics[width=0.32\linewidth,height=0.35\linewidth]{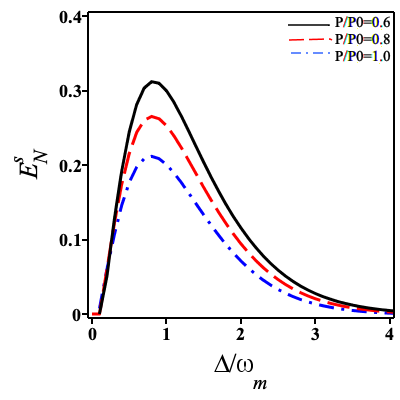}
\includegraphics[width=0.32\linewidth,height=0.35\linewidth]{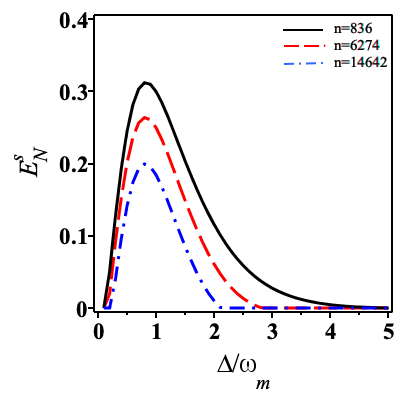}
\includegraphics[width=0.32\linewidth,height=0.35\linewidth]{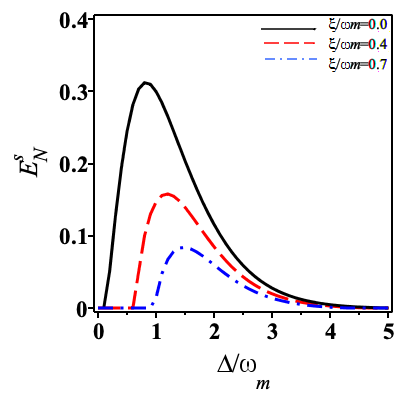}
\caption{(Color online)The steady-state logarithmic  negativity $E^s_{N}$ versus the normalized modified detuning  $\Delta/\omega_m$, (a) for different values of the  input field power $P$, (b) for different values of the mean number of thermal phonons $\bar{n}$ , i.e., for different values of temperature $T$, where $\xi=0$. (c) for different values of the photon hopping strength $\xi$, where $\bar{n}=836 (T=0.4K)$ and $P=P_0=50 \emph{mW}$.  The parameters are chosen to be ($L=1\;  \emph{mm}, m_{j} = 5\; \emph{ng} , \omega_{m}/2\pi = 10\;  \emph{MHz}, \gamma_{m}/2\pi =100\;  \emph{Hz}, \kappa/2\pi = 14\;  \emph{MHz} , \lambda =810\; \emph{nm}, N=0, M=0$).}\label{Fig3}
\end{figure}

Firstly, we assume that the two cavities are coupled only via the photon hopping process, which can be established between the two optomechanical cavities by a waveguide and exploring the bipartite entanglements in the proposed scheme. Here, we ignore the squeezed driven light and examine the steady state entanglement of the possible mutual subsystems in terms of the logarithmic negativity $E_N$. We note the logarithmic negativities for the cavity 1-mirror 1, cavity 2-mirror 2, cavity 1-cavity 2 and mirror1-mirror 2 as $E_N^{F_1M_1} , E_N^{F_2M_2},  E_N^{F_1F_2} \textrm{and }E_N^{M1M2}$, respectively. Moreover, we assume that the two cavities are identical and the effective cavity detuning satisfies $(\Delta_1=\Delta_2=-\Delta)$. In this case, the only entanglement that can be generated is between the resonator and intracavity mode in each optomechanical cavity. The photon hopping process between optomechanical cavities can not transfer entanglement between different bipartite systems. Thus, the logarithmic negativity between optical modes and mechanical resonators are null within the considered parameter regimes and with the proposed scheme. Meanwhile, the entanglement inside each coupled optomechanical cavity can be generated and affected with a variation of strength photon hopping process. Indeed, we are concerned with the resonator-intracavity mode entanglement, which is described by the logarithmic negativity $E_N^{F_1M_1} = E_N^{F_2M_2}=E_{N}^s$. We present in Fig. \ref{Fig3}(a, b) the logarithmic negativity between the cavity j- mirror j $E_N^{s}$  in terms of the normalized detuning $\Delta/\omega_m$  for the case where the two cavities are not coupled and the optical modes are in the vacuum state $N=0, M=0$, for different values relative power and the mean number of thermal phonons $\bar{n}$ . We recover here the same previous results, where the entanglement increases with the normalized detuning until a certain maximum of around $\Delta/\omega_m\approx 1$, then it decreases. Also, the entanglement increases with the power of the input light as it is shown in Fig. \ref{Fig3}(a). On the other hand, when the mean number of phonons is increased the entanglement decreases and can survive for large temperatures, i.e., the entanglement of the resonator with the intracavity mode is very robust against the temperature \cite{Rehaily2017,Genesc,Vitali2007} as it is shown in Fig. \ref{Fig3}(b).
When the photon hopping process is established between the optomechanical cavity, we can see from Fig. \ref{Fig3}(c) that the entanglement of the resonator and the intracavity decreases with increasing the photon hopping strength and the maximum will be shifted to largely normalized detuning. This can be explained in Fig. \ref{Fig2}(b) when the photon hopping process is established and increased, the correlated photons with resonators are transmitted to the waveguide, thus $ |a^s_{j}|$ decreases and the effective coupling rate $G_j$ decreases, thus the degree of entanglement between the resonator and the intracavity mode decreases. 

\begin{figure}[ht]
\center{
\hspace*{1cm}\textbf{(a)} \hspace*{6.cm}\textbf{(b)}\\
\includegraphics[width=0.45\linewidth,height=0.4\linewidth]{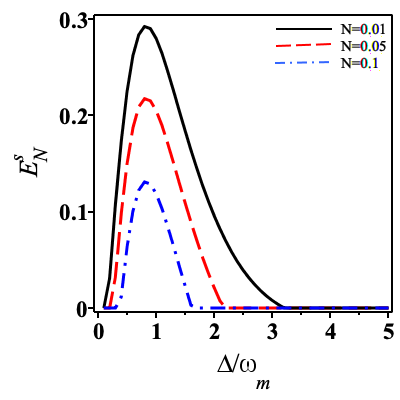}
\includegraphics[width=0.45\linewidth,height=0.4\linewidth]{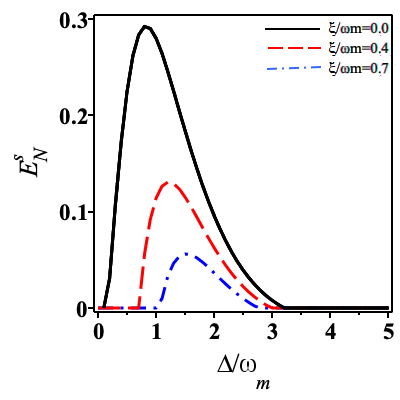}}
\caption{(Color online)The steady-state logarithmic  negativity $E^s_{N}$ versus the normalized modified detuning  $\Delta/\omega_m$, (a) for different values of the  mean number of photons $N$, where $M=\sqrt{N(N+1)}$ and $\xi=0$.  (b) for different values of the photon hopping strength $\xi$, where $\bar{n}=836$ and $N=0.01$. The parameters are chosen to be ($L=1\;  \emph{mm}, m_{j} = 5\; \emph{ng} , \omega_{m}/2\pi = 10\;  \emph{MHz}, \gamma_{m}/2\pi =100\;  \emph{Hz}, \kappa/2\pi = 14\;  \emph{MHz} , \lambda =810\; \emph{nm}.$) }\label{Fig4}
\end{figure}

\begin{figure}[h]
\center{
\hspace*{0.cm}\textbf{(a)} \hspace*{5.cm}\textbf{(b)}\\
\includegraphics[width=0.45\linewidth,height=0.33\linewidth]{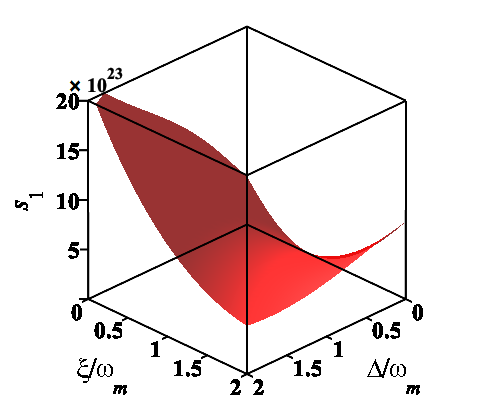}
\includegraphics[width=0.45\linewidth,height=0.33\linewidth]{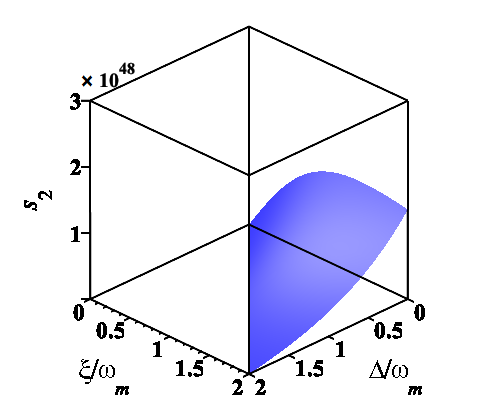}}
\caption{(Color online)The stability conditions in the red-detuning region versus the normalized modified detuning  $\Delta/\omega_m$  and the normalized photon hopping $\xi/\omega_m$. (a) $s_1(\Delta, \xi)$ and   (b) $s_2(\Delta, \xi)$. The parameters are chosen as in previous figure. }\label{Fig5}
\end{figure}

Secondly, now we assume that the two optomechanical cavities are couples via a squeezed light source (SLS), which couples the single optical modes. We explore the entanglement between the resonator and intracavity mode in each optomechanical cavity. As the two coupled cavities are identical, we have $E_N^{F_1M_1} = E_N^{F_2M_2}=E_{N}^s$.  In the absence of the photon hopping process, we plot in Fig. \ref{Fig4}(a),  the logarithmic negativity between the resonator and the intracavity mode, in terms of the normalized detuning  and for different values of the squeezed degree. The robustness of such a mirror-optical mode ensemble entanglement concerning thermal environment of optical mode is presented in Fig. \ref{Fig4}(a). As obviously shown in Fig. \ref{Fig4}(a), owing to thermal environment-induced decoherence, the intensity of the mirror-optical mode ensemble entanglement decreases and eventually vanishes with the augmentation of thermal environmental temperature. The critical normalized detuning that corresponds to $E^{s}_N=0$ decreases with increasing the average number of photons in the optical mode $N$.  Also, we note that the maximum of entanglement remains always around $\Delta/\Omega_m \approx 1$ when $\xi=0$. This can be understood when $N$ is increased, each cavity mode becomes in the thermal environment and it is well known that the entanglement decreases \cite{Genesc,Rehaily2017,Bai2016,Sete2015,Kuzyk2013,Zhang,Sete:14}. 
With the increase of coupling strength $\xi$, the entanglement in thermal squeezed reservoir decreases and its maximum is shifted to large values of normalized detuning. Further, as clearly shown in Fig. \ref{Fig4}{(b)}, the critical values of the normalized detuning, where the entanglement vanishes, increases and are shifted away from one $(\Delta_c \simeq \Delta +\xi)$. Indeed, for identical optomechanical cavities, we have $\kappa_1=\kappa_2=\kappa$, $G_1=G_2=G$, and $\Delta_1=\Delta_2=\Delta$.  Then the Eqs. \ref{eq8} can be written as 

\begin{eqnarray}\label{eq24}
\delta\dot{Q}&=&\omega_{m} (\delta P), \nonumber\\
 \delta\dot{P}&=&-\omega_{m}  \delta Q-\gamma_{m}  \delta P+ G \delta X +F_1, \nonumber\\
    \delta\dot{X}&=&-\kappa \delta X - (\Delta+\xi ) \delta Y+ F_2,      \nonumber\\
 \delta\dot{Y}&=&-\kappa \delta Y+ (\Delta +\xi )\delta X+ G \delta Q + F_3,
 \end{eqnarray}
where  $\delta Q= \delta q_1+ \delta q_2$, $\delta P= \delta p_1+ \delta p_2$, $\delta X=\delta X_1+ \delta X_2$ and $\delta Y=\delta Y_1+ \delta Y_2$. Eqs. \ref{eq24} are the quantum Langevin equations of one optomechanical cavity with an effective detuning $\Delta+\xi$, where $F_1$ is the quantum brownian  stochastic force with zero mean value, while $F_{1(2)}$ are the input noise operators, which satisfy the correlations relations Eqs.(\ref{eq3}, \ref{eq5}). Furthermore, it has shown \cite{Vitali2007,vitali2007entangling,Genesc,Genes2009,Barzanjeh,nunnenkamp2011single} that the logarithmic negativity of the system is maximum around the effective detuning  $\Delta'=\Delta +\xi \approx -\omega_m$, which gives an explanation of the shift of the maximum of entanglement to high effective detuning as it is shown in Fig.\ref{Fig4}(b). On the other hand, the stability analysis can be performed on the linearized set of Eqs. (\ref{eq24}) by using the Routh-Hurwitz criterion \cite{Gradshteyn2014}. Two conditions are obtained
\begin{eqnarray}\label{eq25}
s_1&=&\omega_m(\kappa^2+\Delta'^2)-G^2\Delta'> 0 \nonumber \\
s_2&=&2\gamma_m\kappa \Big\{\Big [\kappa^2+(\omega_m-\Delta')^2\Big]\Big[\kappa^2+(\omega_m+\Delta')^2 \Big] 
+ \gamma_m\Big[(\gamma_m+2\kappa)(\kappa^2+\Delta'^2)+2\kappa\omega_{m}^2\Big]\Big\}, \nonumber\\ 
&+&\Delta'\omega_m G^2(\gamma_m+2\kappa)^2 > 0.
 \end{eqnarray}
The violation of the first condition $s_1< 0$ shows the rise of the well-recognized effect of bistable character observed in \cite{Dorsel1983, Gozzini1985} and arises only for positive detuning
$(\Delta’ > 0)$. The violation of the second condition, $s_2 < 0$, specifies instability in the domain
of a blue-detuned laser $(\Delta’ < 0)$ and it matches to the appearance of a self-sustained oscillation regime where the resonator effective damping rate
becomes zero. In this regime, the laser field energy transmits into field harmonics at frequencies $\omega_L \pm \beta \omega_m,  (\beta= 1, 2,. . .)$ and also provides the resonator coherent oscillations. A complex multi-stable regime can arise as explained in \cite{marquardt2006dynamical}. 
In Fig. \ref{Fig5} we plot the two conditions in the red-detuning region, where the conditions are simultaneously satisfied for $\xi \geq \Delta$.

\begin{figure}[ht]
\hspace*{1.5cm}\textbf{(a)} \hspace*{3.5cm}\textbf{(b)}\hspace*{3.5cm}\textbf{(c)}\\
\includegraphics[width=0.32\linewidth,height=0.33\linewidth]{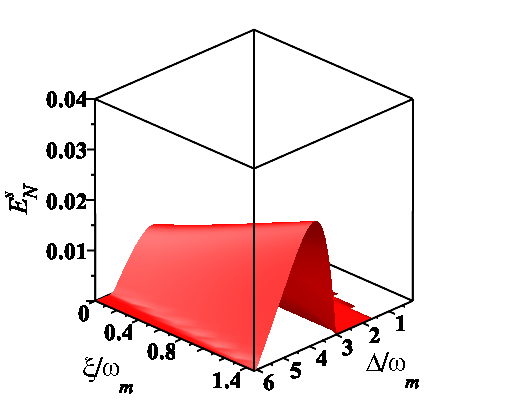}
\includegraphics[width=0.32\linewidth,height=0.33\linewidth]{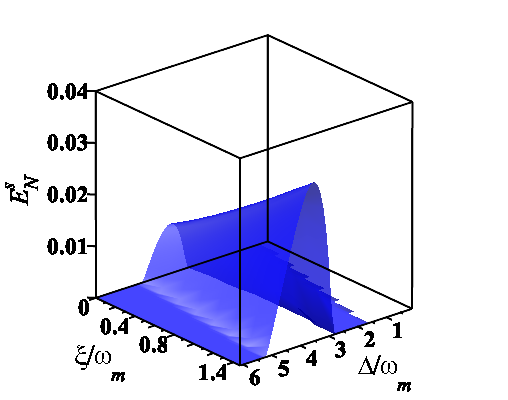}
\includegraphics[width=0.32\linewidth,height=0.33\linewidth]{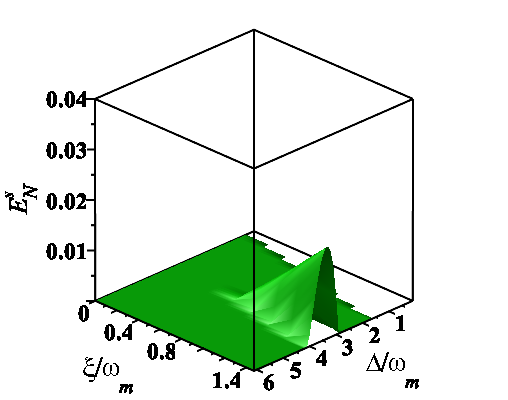}
\caption{(Color online)The steady-state logarithmic  negativity $E^s_{N}$ versus the normalized modified detuning  $\Delta/\omega_m$ and the photon hopping strength $\xi/\omega_m$,  (a) for  $N=0.025$, (b) for  $N=0.05$ and (c) for  $N=0.1$, where in all cases  $M=\sqrt{N(N+1)}$, where $\bar{n}=836$ and $P=P_0=50 \emph{mW}$.  The parameters are chosen to be ($L=1 \emph{mm}, m_{j} = 5\emph{ng} , \omega_{m}/2\pi = 10\;  \emph{MHz}, \gamma_{m}/2\pi =100 \; \emph{Hz}, \kappa/2\pi = 14\;  \emph{MHz} , \lambda =810\; \emph{nm}$).}\label{Fig6}
\end{figure}

On the other hand, we now investigate the generation of the stationary entanglement between the optical modes, through the logarithmic negativity $E_N^{F_1F_2} =E_{N}^s$, when the two cavities are coupled by the squeezed light source. With our dispositive and within the experimental parameters, the bipartite entanglements between the cavity 1-mirror 2, cavity 2-mirror 1 ensemble, and mirror1- mirror 2  are so weak that there is no need to consider them. In meanwhile, the entanglement between cavity 1- cavity 2 ensemble can be considered and it is originated from the squeezing effect between the optical modes. The consequences on the behavior of this bipartite entanglement are exhibited in Fig. \ref{Fig6}. In which we plot the bipartite logarithmic negativity $E_{n}^s$ in terms of the normalized detuning and the normalized photon hopping strength for different values of the thermal squeezed parameters $N$ and $M=\sqrt{N(N+1)}$ at a fixed temperature of $\bar{n}=836$ and same previous parameters. It is obvious that there is a generation of kind of entanglement between the optical mode of each optomechanical cavity, which is essentially due to the squeezed light that established between the two separated cavities. In addition, it is clear that there is a link of entanglement transfer among the three bipartite subsystems, i.e., field 1-mirror 1, field 2-mirror 2 and field 1-filed 2. Indeed, the bipartite entanglements $E^{F1-F2}$ increases while the bipartite entanglements $E^{F1-M1}$ and $E^{F2-M2}$  decreases with the increase of thermal squeezing parameters. The maximum of the entanglement between optical modes is reached at large values of normalized detuning and for $N\approx 0.05$ with $M=\sqrt{N(N+1)}$. It is notable that, consequently, the redistribution of entanglement between the three bipartite subsystems is predominant when the thermal squeezed parameters are well appropriate with the Stocks sideband.

The transfer of squeezing light to the vibrating membrane or a movable mirror in an optomechanical system is an interesting subject and well investigated \cite{Jaehne2009}. Indeed, an optical cavity is driven by the squeezed light and couples via the radiation pressure to the membrane or mirror, effectively providing a squeezed heat bath for the mechanical oscillator. Thus, we have proposed a scheme of two optomechanical cavities, then we can generate a two-mode squeezed heat bath for the mechanical oscillators. Besides, it has shown that the two-mode squeezed vacuum state can be used for the teleportation of an unknown quantum state \cite{milburn1999quantum}. However, restrictions can result in microscopic scales.

On the other hand, optomechanical systems already are a quantum hybrid system, which couple two quantum systems of a different physical nature: light and mechanical oscillations. In general, hybrid approaches are useful for purposes such as quantum information processing, in order to merge the perfections of diverse physical systems in one structure. Besides the optomechanical systems push the target of quantum teleportation to the largest object (macroscopic scales) so far. 

With the recent development of optomechanics, it has shown that quantum teleportation from light beams to vibrational states of a macroscopic diamond under ambient conditions can be experimentally realized with average teleportation fidelity exceeding the threshold value for secure teleportation \cite{Hou2016a}. Further, the optomechanical systems can be used to link microwave and optical frequency and therefore can be used for making telecom quantum state transfer \cite{Rueda2019}.

Thus, to analyze the entanglement of a traveling CM bipartite Gaussian system, constituted of the categorized output optical modes $(a_1)$ and $(a_2)$, the covariance matrix $W_{a_1a_2}$ of the reduced Gaussian state $\hat{\rho}_{a_1a_2}$ can be achieved by removing the mechanical mode, i.e., by eliminating the rows and columns of the covariance matrix W consistent to the latter mode. The reduced covariance matrix is given by Eq. (\ref{eq20}) where the diagonal blocks are the covariance matrices corresponding to the optical modes $(a_1)$ and $(a_2)$, whereas the upper non-diagonal block defines the correlation between the optical modes. 

Once the two traveling optical output fields are entangled, they can be manipulated for quantum teleportation of an unknown coherent state \cite{Braunstein}. For long-distance purposes, it is essential to take the robustness of the resulting quantum communication channel with matter to optical losses, which are inevitable while the two fields travel a long distance in free space or down an optical fiber. Losses can be explained using a beam-splitter model \cite{Barbosa}. 

\begin{figure}[h]
\includegraphics[width=0.65\linewidth,height=0.5\linewidth]{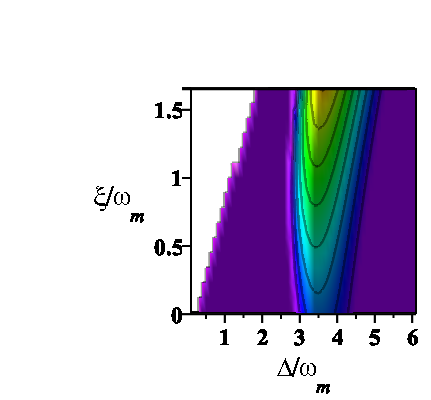}
\includegraphics[width=0.3\linewidth,height=0.5\linewidth]{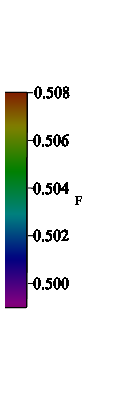}
\caption{(Color online)The fidelity of coherent state of optical modes versus the normalized detuning and photon hopping strength, for $N=0.05$ and $M=\sqrt{N(N+1)}$. The other parameters are as in previous figure. The solid curves correspond just to the conditions of stability. }\label{Fig7}
\end{figure}

In fact, if the input state is a single-mode Gaussian state with covariance matrix $W_{in}$, the fidelity of teleportation reads \cite{Barzanjeh,Barzanjeha}
\begin{equation}
\label{ eq22}
\mathcal{F} = \frac{2}{\sqrt{Det(2W_{in}+Z)}},
\end{equation}
where  $Z= SW_1S+SW_c +W^T_{c}S+W_2$, with $S$ is the diagonal Pauli matrix $S=diag(1, -1)$. The $2 \times 2$ matrix N is semipositive definite, $N \geq 0$, it expresses the noise added to the teleported state, and it is equal to zero only in the ideal situation of perfect EPR correlations between the output optical modes, where $W_{in}$ is the covariance matrix of the teleported Gaussian state and $W_1, W_2$ and $W_c$ are the matrices in Eq. (\ref{eq20}) in the presence of optical loss, while in its absence they would be the same as in Eq. (\ref{eq20}). We are continuously concerned with an input coherent state where $W_{in} = \textit{I}$, where $ \textit{I}$ is the $2 \times 2$ identity matrix. 
Furthermore, the fidelity concerning the optimal upper bound, described in \cite{Mari} and obtained by optimizing overall probable local operations, is given by 
\begin{equation}
\label{ eq23}
\mathcal{F} = \frac{1}{1+e^{-E_N}},
\end{equation}
where $E_N$ is the logarithmic negativity of the quantum channel. In Fig.\ref{Fig7}, we plot the fidelity of coherent state of the optical modes in terms of the normalized detuning and the photon hopping strength for the optimized values of the thermal squeezed environment, which are the same for entanglement.

\section{Conclusions}\label{sec6}
In conclusion, we have suggested a useful scheme to generate robust entanglement between different subsystems within the experimental parameters. Moreover, we have shown that the photon hopping process can affect the resonator-intracavity mode entanglement. This is due to the transmission of the correlated photon to the waveguide. On the other hand, when the two optomechanical cavities are driven by a squeezed light source, vigorous optical modes entanglement can be established. cavities are driven by a squeezed light source, vigorous optical modes entanglement can be established. In this case, a repartition of bipartite entanglement is achieved and it has shown that the decrease of the resonator and intracavity mode entanglement corresponds to an increase of entanglement between optical modes.
Using the experimentally reachable parameters, we have provided an important limit on the photon hopping strength and the values of the thermal squeezed light for robust bipartite entanglement in the proposed scheme. \\ 
It has also shown that the stationary bipartite entanglement persists for an appropriate value of temperatures. The fidelity of coherent states of optical modes is explored and offered interesting conditions on the stability of the system.  The stability conditions are the same for robust entanglement generation. Finally, we have remarked that the entanglement transfer based on this coupled system can be recognized. Such a hybrid system can be used for the realization of quantum memories for continuous variable quantum information processing and quantum-limited displacement measurements.

\begin{acknowledgements}
The researchers acknowledge the deanship of Scientific Research at Imam Mohammad Ibn Saud Islamic University (IMSIU), Saudi Arabia, for financing this project under grant no. (381213)
\end{acknowledgements}

\bibliographystyle{spphys}  
\bibliography{Ref1}
\end{document}